\documentclass[aps,paper,12pt]{revtex4-2}
\usepackage{amsmath}
\usepackage{graphicx}
\usepackage{dcolumn}
\usepackage{bm}
\usepackage{epsfig}
\usepackage{graphics}
\usepackage{rotating}
\usepackage{xcolor}
\graphicspath{{./figures/}}
\usepackage{float}
\begin{document}


\title{Gravitational particle production, the cosmological tensions and fast radio bursts}
\author{Recai Erdem}
\email{recaierdem@iyte.edu.tr}
\affiliation{Department of Physics \\
{\.{I}}zmir Institute of Technology\\
G{\"{u}}lbah{\c{c}}e, Urla 35430, {\.{I}}zmir, T{\"{u}}rkiye}


\begin{abstract}
In \cite{Erdem-Universe} it had been found that gravitational particle production (to be more specific, gravitational vacuum polarization) results in an effective increase in the directly measured value of the Hubble constant $H_0$ while it does not affect the value of the Hubble constant derived from energy densities $\bar{H}_0$. It had also been pointed out that this may explain why the Hubble constant $H_0$ determined from direct measurements (such as in SN Ia measurements)  and the Hubble constant  determined from indirect measurements (such as in CMB calculations in the framework of $\Lambda$CDM) are different. In the present study, first I correct a misidentification in \cite{Erdem-Universe}, namely,  $\hat{H}_0=\left(\frac{\bar{H}_0}{H_0}\right)\bar{H}_0$  (rather than $\bar{H}_0$) is the value of the Hubble constant measured in CMB and BAO measurements. Then I extend the analysis to the $\sigma_8$ tension, and to determination of the Hubble constant through observations of fast radio bursts. It is observed that inclusion of the effect of gravitational vacuum polarization essentially does not neither mitigate nor exacerbate the $\sigma_8$ tension (while it mitigates or relieves the Hubble tension). This result is significant in the light of the studies in literature that question existence of a true $\sigma_8$ tension. Moreover, the present framework predicts that the value of the Hubble constant measured in fast radio bursts is $\hat{H}_0$ as in CMB and BAO measurements. This may be checked with observations in future after more precise and conclusive measurements of $\hat{H}_0$, $\bar{H}_0$, $H_0$.
\end{abstract}

\maketitle

\section{introduction}

 Clustering of matter distribution on scales of 8 $h^{-1}$ Mpc may be specified by the amplitude of matter fluctuations $\sigma_8$  or the derived quantity $S_8$ = $\sigma_8\,\sqrt{\frac{\Omega_m}{0.3}}$ (which is the more suitable parameter in cosmic shear measurements). The $S_8$ or $\sigma_8$ tension is the discrepancy between their values directly measured at low redshifts (such as that of KiDS -1000) and the ones measured by CMB measurements (such as that of Planck) and projected to the present day under the assumption of the applicability of $\Lambda$CDM for evolution of the perturbations \cite{amplitude,S8,review-status}. For example, KiDS-1000 collaboration found $S_8$ = $0.759\,^{+0.024}_{-0.021}$ \cite{KiDS-1000} while the Planck collaboration found $S_8$ = $0.834\,\pm\,0.016$ \cite{Planck}. These measurements are in 2.6$\sigma$ tension. This discrepancy may be identified as $\sigma_8$ tension once matter density parameter $\Omega_m$ is assumed to be fixed. Given the strong constraints on the value of the matter density parameter $\Omega_m$ obtained by Planck this result may imply a tension in $\sigma_8$ in Planck measurements \cite{Planck}. However, such a direct correspondence is not verified in general. For example, DESI finds a possible tension between the values of $\Omega_M$ determined in different experiments \cite{DESI1,DESI2}. In the present paper we will mainly consider the $\sigma_8$ part of the tension.

A severer cosmological tension is the Hubble tension. Hubble tension is the discrepancy between the direct local measurements of Hubble constant by type SN Ia supernovas calibrated by Cepheids \cite{Riess} and the measurements of Planck \cite{Planck} and other indirect measurements such as baryon acoustic oscillations (BAO) \cite{BAO1,BAO} that make use of the assumption of the applicability of $\Lambda$CDM before and after recombination. The values of Hubble constant obtained from direct measurements are higher than the indirect ones. For example, \cite{Riess} finds the Hubble constant as $\left(73.04\pm\,1.04\right)km\,s^{-1}\,Mpc^{-1}$ while \cite{Planck} finds it as $\left(67.4\pm\,0.5\right)km\,s^{-1}\,Mpc^{-1}$. These measurements are in 4.9$\sigma$ tension. SN Ia supernova and Planck measurements may differ by more than 5$\sigma$ \cite{tension,Kamionkowski,Dainotti,Khalife}.

There are many theoretical studies that aim to resolve or mitigate the Hubble tension \cite{tension,Kamionkowski,Dainotti,tension2,Dainotti2,S8,why} or the $\sigma_8$ tension \cite{S8,visual}. However, the theoretical models that aim to ease the Hubble tension generically exacerbate the $\sigma_8$ tension \cite{why,worsen}. It seems rather difficult to resolve both of the Hubble and $\sigma_8$ tensions simultaneously \cite{generic,simultaneous,visual}.

Gravitational particle production (GPP) \cite{Imamura}, is a generic property of quantum field theory in Friedman-Lema{\^{i}tre-Robertson-Walker (FLRW) spacetimes \cite{Parker,QFTC1,QFTC2,Ford,Kolb,without}. In \cite{Erdem-Universe}, I had shown that GPP (or to be more concrete, gravitational vacuum polarization (GVP), which is a specific manifestation of gravitational particle production for small values of Hubble parameter) increases the directly measured value of the Hubble constant while it leaves energy densities intact, so leaves the value of the Hubble constant obtained from energy densities the same. This may explain the difference between the values of the Hubble constant, namely, the one\textcolor{blue}{s} in direct measurements such as in SN Ia measurements and the one\textcolor{blue}{s} in the indirect CMB and BAO measurements if the total mass of the scalars in the related particle physics model is sufficiently high.

Fast radio bursts (FRB) are millisecond duration radio frequency electromagnetic transients of extra-galactic origin \cite{FRB0,FRB1,FRB2,FRB3}. FRB observations may be related to Hubble constant through dispersion measure (DM) \cite{FRB-Hubble1,FRB-Hubble2,FRB-Hubble3}. Hence, FRB observations are used as an alternative method to determine the Hubble constant  \cite{Hagstotz,Wu,Liu,DM-MW,Zhao,Wei,Hoffmann,Yang,Gao,Wang,Piratova-Moreno}.

In the following, in Section II., first the main lines of the impact of GVP on the Hubble tension \cite{Erdem-Universe}, especially emergence of two values of the Hubble constant, namely, the one in direct measurements and the one employed in parametrization of energy densities, are briefly summarized since this is one of the basic elements of the discussion in the following parts. Then, the impact of GVP on CMB and FRB measurements of the Hubble constant are studied. In Section III. the main aspects of the impact of GVP on direct measurements of the Hubble constant (i.e. $H_0$) and its impact on CMB and FRB measurments (i.e. $\hat{H}_0$) are discussed. In Section IV, the effect of GVP on $\sigma_8$ is discussed. Finally, Section V. summarizes the main conclusions.

\section{Impact of gravitational vacuum polarization on the Hubble tension: A brief overview}

\subsection{Effect of gravitational vacuum polarization on Hubble constant}

In this subsection we briefly mention the basic features of gravitational vacuum polarization (which is a specific manifestation of  gravitational particle production) and its impact on directly measured value of Hubble constant that is discussed in \cite{Erdem-Universe} (and summarized in Appndix A).
Gravitational particle production (GPP) is a generic property of quantum field theory in curved spaces with time dependent backgrounds, so, GPP generically takes place in cosmological backgrounds \cite{Parker,QFTC1,QFTC2,Ford,Starobinsky,book,Grib2,Chung,fermion,vector,Grib,Boyanovsky}. The number and energy densities of the particles that are produced by GPP may be calculated in a straightforward way (although it may be difficult to obtain \textcolor{blue}{it} in general). In particular in \cite{Erdem-Universe} a scalar field $\phi$ of mass $m_\phi$ in the background of a spatially flat Robertson-Walker spacetime is considered, and it is assumed that the universe is described by $\Lambda$CDM or a model that (approximately) mimics $\Lambda$CDM.
In such a background the energy density of the gravitationally produced $\phi$s after decoupling may be approximated by \cite{Starobinsky,Erdem-Universe}
\begin{equation}
\rho^{(PP)}
\,=\,\frac{\hbar}{96\,\pi\,c}\,
\left(\frac{m_\phi\,c}{\hbar}\right)^2H^2\;,
\label{eq:48}
\end{equation}
where $H$ is the Hubble parameter. (see Appendix A for a brief overview of the derivation of (\ref{eq:48})),

The total effective energy density $\rho^{eff}$ (that is the sum of the energy density of background $\rho^{(bg)}$ and the energy density of gravitationally produced particles $\rho^{(PP)}$) relates to Hubble parameter $H$ as
\begin{eqnarray}
\frac{3\,c^2}{8\pi\,G}\,H^2&&=\,\rho^{eff}\,=\,\rho^{(PP)}\,+\,\rho^{(bg)}
\,=\,\frac{\hbar}{96\,\pi\,c}\,\sum_i\left(\frac{m_i\,c}{\hbar}\right)^2\,H^2
\,+\,\rho^{(bg)}
\nonumber\\
&\simeq&\rho^{(bg)}
\,+\,1.8\,\times\,10^{-58}\times\,
\sum_i\left(\frac{m_i\,c^2}{eV}\right)^2\,
\frac{3\,c^2}{8\pi\,G}\,H^2
\label{eq:49c} \\
&&\mbox{i.e.}~~~3\,H^2\,=\,\frac{8\pi}{c^2}\,\left[\frac{G}{1\,-\,1.8\,\times\,10^{-58}\times\,
\sum_i\left(\frac{m_i\,c^2}{eV}\right)^2}\right]\,\rho^{(bg)} \;, \label{eq:49d}
\end{eqnarray}
where $\rho^{(bg)}$ is identified as the total energy density (of the background), and $m_i$ denotes the mass of the $ith$ scalar particle (that contributes to $\rho^{(PP)}$). 

The essence of Eq.(\ref{eq:49d}) may be seen more clearly by re-expressing it as
	\begin{eqnarray}
		3\,H^2&=&\frac{8\pi}{c^2}\,G\left(\rho^{(bg)}\,+\,\rho^{(PP)}\right)\,=\,\frac{8\pi}{c^2}G_{eff}\;\rho^{(bg)} \;, \label{eq:49ddd}
\end{eqnarray}
where $G_{eff}$ is 
\begin{equation}
	G_{eff}\,=\,\frac{G}{1\,-\,1.8\,\times\,10^{-58}\times\,\sum_i\left(\frac{m_i\,c^2}{eV}\right)^2}\;,
	\label{s8-eq:2a}
\end{equation}
 Note that the rescaling of the gravitational constant in (\ref{s8-eq:2a}) is valid for all redshifts in the measurements of the Hubble constant that directly use the expansion rate of the universe by employing the fundamental Friedmann equation (on contrary to the models with a sudden transition in its value from a constant value to another constant value). However, the applicability of (\ref{s8-eq:2a}) for times before photon decoupling is not guaranteed since the approximate formula given in Eq.(\ref{eq:49d}) is derived for the times after decoupling when the strong form of the adiabatic condition  (\ref{eq:27}) is applicable.
 
 There are two crucial observations above. The first is that $\rho^{(PP)}$ in (\ref{eq:48}) is proportional to $H^2$. This makes it possible to include the effect of GVP as a re-scaling of the gravitational constant as in (\ref{s8-eq:2a}).  
The other crucial observation is the following. The masses of all standard particles (such as baryonic matter, radiation) are very small with respect to the Planck mass. We assume that the same is true for all dark matter and  dark energy particles i.e. they are made of relatively light particles (if they consist of elementary particles).  Further, we assume that there are initially no appreciable amounts of particles with the Planck scale masses (such as $\phi$ particles) in the background i.e. their effect is almost wholly through GVP as in Eq.(\ref{eq:49d}). Hence, Eq.(\ref{eq:49d}) (and similar expressions for fermions \cite{fermion} and vector particles \cite{vector}) tells us that GVP does not give appreciable amount of additional contribution to $\rho^{(bg)}$ i.e. GVP essentially does not change the energy densities of baryonic matter, dark matter, dark energy, radiation. Therefore, GVP may be considered effectively to re-scale the value of the gravitational constant (through Eqs. (\ref{eq:49c}) and (\ref{eq:49d})). Note that the effective energy density of $\phi$ particles induced by GVP cannot be considered as a contribution to matter or radiation or dark energy since it is evident from (\ref{eq:48}) that it is proportional to $H^2$, so tracks its evolution. 

Eq.(\ref{eq:49d}) implies that gravitational particle production has a significant contribution to the effective Hubble parameter if $\sum_i\left(\frac{m_i\,c^2}{eV}\right)^2$ is not extremely smaller than $10^{58}$. Note that, ultra-heavy particles are extensively studied in literature \cite{Chung,Grib}. Increasing the mass of $\phi$ results in an overall increase in the total energy density irrespective of the masses and the ratio of the particles in the background. Therefore, the effective increase in the energy density in (\ref{eq:49d}) may be considered as an effective increase in the Newton's gravitational constant $G$. This mechanism is analogous to vacuum polarization in quantum electrodynamics (QED). Vacuum polarization in QED (after renormalization) causes an effective re-scaling in the electromagnetic coupling constant due to pairs of electrically charged virtual pairs.  Therefore, the gravitational particle in this context obtained may be identified as some sort of gravitational vacuum polarization \cite{Starobinsky, QFTC2}. In other words, the phenomena that results in Eq.(\ref{eq:49d}) may be considered as the gravitational vacuum polarization (GVP) manifestation of GPP. (A more rigorous discussion of this point may be found in Appendix B).

It is evident from (\ref{eq:49d}) that gravitational production of $\phi$ particles results in an effective overall increase in the value of Hubble parameter, hence in the value of Hubble constant.
In other words, the effective increase of $G$ in (\ref{eq:49d}) causes an increase in the overall value of the Hubble parameter,
 \begin{equation}
H^2\,=\,\left[\frac{1}{1\,-\,1.8\,\times\,10^{-58}\times\,
\sum_i\left(\frac{m_i\,c^2}{eV}\right)^2}\right]\,\bar{H}^2 \;, \label{eq:49da}
\end{equation}
so an increase in Hubble constant, namely,
 \begin{equation}
H_0^2\,=\,\left[\frac{1}{1\,-\,1.8\,\times\,10^{-58}\times\,
\sum_i\left(\frac{m_i\,c^2}{eV}\right)^2}\right]\,\bar{H}_0^2 \;, \label{eq:49daa}
\end{equation}
where the subscript $0$ stands for the present time, and $\bar{H}_0\,=\,\sqrt{\frac{8\pi\,G}{3}\,\rho_0}$ is the value of the Hubble constant without the effect of GVP included while $H_0$ is the value of the Hubble constant after inclusion of the effect of GVP. Eq.(\ref{eq:49d}) implies that $H_0$ is the value of the Hubble constant determined in direct measurements (through expansion rate of the Universe). 

On the other hand, as we have mentioned above, the energy densities of relatively light particles are not affected by GVP significantly, so $\rho^{(bg)}$ essentially remains the same after inclusion of the effect of GVP. Therefore,
\begin{equation}
\rho^{(bg)}\,=\,\frac{3H^2}{\frac{8\pi}{c^2}\,G_{eff}}\,=\,\frac{3\bar{H}^2}{\frac{8\pi}{c^2}\,G}\;,
\label{s8-eq:0}
\end{equation}
 essentially is not affected by GVP (where $H$ is related to $\bar{H}$ by (\ref{eq:49da})). In other words, $\rho^{(bg)}$ is almost the same before and after the inclusion of the effect of GVP and it may be identified as the total energy density of the universe, and the ratio of its components also remain the same.

\subsection{The value of the Hubble constant in CMB and BAO measurements}

 Hubble constant may be also determined from the imprints of baryon acoustic oscillations on CMB or on large scale structure anisotropies by measuring the angle $\theta$ subtended by sound horizon
\begin{equation}
\theta\,=\,\frac{r_s}{D_A} \;, \label{eq:54}
\end{equation}
where $r_s$ is the comoving size of the sound horizon, $D_A$ is the comoving angular diameter distance to the observed position \cite{Kamionkowski,BAO}. If $\Lambda$CDM is applicable both before and after photon decoupling, then
\begin{equation}
r_s\,=\,\int_{z_a}^\infty\,\frac{v_s(z)\,dz}{H_0\,E(z)}~,~~~~
D_A\,=\,c\int_0^{z_b}\,\frac{dz}{H_0\,E(z)} \;,
\label{eq:54a}
\end{equation}
where  $z$ denotes redshift; $c$ is the speed of light; $v_s(z)$ is the speed of the sound waves in baryon-photon fluid; $a$ = * or $d$ stand for recombination or drag epoch (for the imprint of the acoustic oscillations on CMB radiation or on galaxy autocorrelation function, repectively); $b$ = * or $obs$ denote the redshifts of recombination or of the observed galaxies; $E(z)\,=\,\sqrt{\Omega_\Lambda\,+\,\Omega_M\,\left(1+z\right)^3\,+\,\Omega_R\,\left(1+z\right)^4}$ with $\Omega_\Lambda$, $\Omega_M$, $\Omega_R$ being the density parameters for cosmological constant, dust, radiation, respectively.

Let us assume (unlike the early or late time solutions of the Hubble tension) that the evolution of the universe before and after the recombination is described by the (unmodified) standard model (i.e. $\Lambda$CDM). (In fact, we have expressed (\ref{eq:54a}) in a form that is more suitable for this case.) It is evident from (\ref{eq:54a}) that $\theta$ in (\ref{eq:54}) is unaffected by the values of $H_0$ in the arguments of $r_s$ and $D_A$. However, the value of the Hubble constant affects $r_s$ and $D_A$ by its effect on $z_a$ by affecting recombination. The effects of a change in $z_a$ on $r_s$ and $D_A$ are not the same since the value of $r_s$ is dominated by the value of $E(z)$ at values of $z$ close to $z_a$ while the value of $D_A$ is dominated by the value of $E(z)$ at values of $z$ close to $z=0$. Hence, a variation of the Hubble constant varies $\theta$ by its effect on $z_a$. Thus, the observational value of the Hubble constant may be determined after finding the best fit values for the Hubble constant and the density parameters corresponding to the observed $\theta$.

Studies of Saha equation and the Boltzmann equation suggest that the value of $z_a$ is determined by the number density of baryons \cite{Erdem-Universe}. This point may be seen explicitly through an analytical formula that expresses $r_d$ in terms of $\Omega_b$ and $\Omega_M$ \cite{two-h}.
A more direct way of observing this is through the following analytic formulas that expresses $z_*$ and $z_d$ in terms of the reduced Hubble constant \cite{Hu1}
\begin{equation}
z_*\,=\,1048\,\left[1+0.00124\,\left(\Omega_b\hat{h}^2\right)^{-0.738}\right]\left[1+g_1\left(\Omega_M\hat{h}^2\right)^{g_2}\right] \;,
\label{eq:58a}
\end{equation}
\begin{equation}
z_d\,=\,1315\,\frac{\left(\Omega_M\hat{h}^2\right)^{0.251}}{1
\,+\,0.659\,\left(\Omega_M\hat{h}^2\right)^{0.828}}\left[1+b_1
\left(\Omega_b\hat{h}^2\right)^{b_2}\right] \;,
\label{eq:58b}
\end{equation}
where $g_1$, $g_2$ are some functions of $\Omega_b\hat{h}^2$, and $b_1$, $b_2$ are some functions of $\Omega_M\hat{h}^2$ whose explicit forms may be found in \cite{Hu1}. Here $h$ in \cite{Hu1} is replaced by $\hat{h}$ that is related to $h$ and $\bar{h}$ by 
\begin{equation}
	\hat{h}\,=\,\left(\frac{\bar{h}}{h}\right)\bar{h}
	\label{eq:58aa}
\end{equation}
This may be seen as follows. 

In the case of recombination, the corresponding Boltzmann equation may be expressed as \cite{Dodelson-book}
\begin{equation}
	a^{-3}\,\frac{{\textcolor{blue}{d}}\left(n_ea^3\right)}{dz}\,=\,\frac{n_e^{(0)}n_p^{(0)}}{H}\,<\sigma\,v>\left\{\frac{n_H}{n_H^{(0)}}-\frac{n_e^2}{n_e^{(0)}n_p^{(0)}
	}\right\}
	\label{eq:58}
\end{equation}
which corresponds to the dominant reaction during the recombination $e^-\,+\,p\,\leftrightarrow\,H\,+\,\gamma$. Here $n^{(0)]}$s are the number densities in thermal equilibrium, $<\sigma\,v>$ is the thermally averaged cross section. The Hubble parameter $H$ in (\ref{eq:58}) results from converting the time derivative in \cite{Dodelson-book} to the derivative with respect to redshift. Eq.(\ref{eq:58}) determine\textcolor{blue}{s} the evolution of $n_e$ as a function redshift, so determines $z_*$ which is the redshift where $n_e$ makes a sharp drop, so Eq.(\ref{eq:58}) implies Eq.(\ref{eq:58aa}) (where $\bar{h}$ and $h$ result from number densities and $H$, respectively). This justifies use of $\hat{h}$ rather than $h$ in (\ref{eq:58a}).

In the case of baryon decoupling, the relevant equation is the Boltzmann equation for peculiar baryon velocities $v_b$ for $k$'th mode (in momentum space) is \cite{Dodelson-book}
\begin{equation}
v_b^\prime\,+\,H\,v_b\,=\,-ik\,\Psi\,+\,\left(\frac{\tau^\prime}{R}\right)\left(v_b+3i\Theta_1\right)
	\label{eq:58c}
\end{equation}
where $\prime$ denotes derivative with respect to conformal time, $\Psi$ is the perturbation to the FLRW metric that corresponds to Newtonian gravitational potential, $R=\frac{3\rho_b^{(0)}}{4\rho_\gamma^{(0)}}$ is the baryon to photon ratio at photon-baryon decoupling, $\Theta_1$ is the dipole moment of photon temperature, and 
\begin{equation}
	\tau(\eta)\,=\, \int_\eta^{\eta_0}d\eta\,n_e(\eta)a
	\label{eq:58ca}
\end{equation}
is the optical depth of CMB photons where $\sigma_T$ is the cross section for Thomson scattering. Hence, (\ref{eq:58c}), in terms of redshift $z$ becomes
\begin{equation}
	\frac{d\,v_b}{dz}\,-\,v_b\,=\,\frac{ik}{H}\Psi\,-\,\left(\frac{n_{e0}\sigma_T}{R}\right)(1+z)^2\left(v_b+3i\Theta_1\right),
	\label{eq:58cb}
\end{equation}
where $n_{e0}$ is the number density of electrons at present.
The coupling of photons to baryons is determined by the last term on the right-hand side of (\ref{eq:58cb}). When this term becomes negligible i.e. at $z_d$ baryons travel as free particles under the effect of gravity. This term is proportional to $\frac{n_{e0}}{H}$, so it is proportional to $\frac{\bar{h}^2}{h}$ i.e. to $\hat{h}$ defined in (\ref{eq:58aa}). This justifies use of $\hat{h}$ rather than $h$ in (\ref{eq:58b}).

  The effect of the energy densities on $z_*$ and $z_d$ (through through their dependence on $\tilde{h}$, $\Omega_b$, $\Omega_M$) is evident in (\ref{eq:58a}) and (\ref{eq:58b}). $z_*$ and $z_d$ are identified as the redshift where $n_e$ makes a dip and $z_d$ as the redshift where the optical depth is 1. (The above equations supplemented by a phenomenological rule observed by \cite{Percival}, namely, in a spatially flat universe $\Omega_M\tilde{h}^{p}$ (where $p=3.4$ in the original paper while $p$ is found to be 3 by Planck) may be determined from the positions of the acoustic peaks while $\Omega_M\tilde{h}^2$ may be directly determined from data analysis for best fists. This information may be used to determine $\Omega_M$, $\tilde{h}$, $\Omega_b$ separately \cite{Planck,Percival}).

\subsection{Gravitational vacuum polarization and fast radio bursts}

Fast radio bursts (FRB) are millisecond duration radio frequency electromagnetic transients of extra-galactic origin \cite{FRB0,FRB1,FRB2,FRB3}. FRB observations may be related to Hubble constant through dispersion measure (DM) \cite{FRB-Hubble1,FRB-Hubble2,FRB-Hubble3}.
DM is defined as the number of electrons in a column of unit cross-sectional area along the line of sight of the observer to the source \cite{Ioka,Inoue,Deng} i.e.
\begin{equation}
DM\,=\,\int_0^d\,\frac{n_e(l)}{1+z_l}\,dl\;,
\label{eq:y1}
\end{equation}
where $dl=c\,dt$, $d$ is the distance of the source to the observer, $z_l$ is the redshift of the (free) electrons at a distance $l$ to the observer, $n_e(l)$ is the electron number density at a distance $l$ to the observer; the $\frac{1}{1+z_l}$ factor is the cosmological correction factor due to expansion of the universe. The observed DM, $DM_{obs}$ is the sum of different contributions \cite{FRB-Hubble2}, namely,
\begin{equation}
DM_{obs}\,=\,DM_{MW}\,+\,DM_{IGM}\,+\,\frac{DM_h}{1\,+\,z_{spec}}\;,
\label{eq:y2}
\end{equation}
where $DM_{MW}$, $DM_{IGM}$, $DM_h$, and $z_{spec}$ are the dispersion measures of the Milky Way, the intergalactic medium between the Milky Way and the host galaxy, the host galaxy of the source, and the spectroscopic redshift of the light coming from the source; respectively.

In particular, $DM_{IGM}$ may be expressed in terms of Hubble constant \cite{FRB-Hubble2,Macquart} by using (\ref{eq:y1}). We repeat the derivation by being careful about the distinction between $H$ and $\bar{H}$ as it will be important in the following discussion. First we observe that $dl\,=\,cdt\,=\,-c\frac{dz}{(1+z)\,H(z)}$. Hence, (\ref{eq:y1}) results in
\begin{equation}
DM_{IGM}
\,=\,\frac{c}{H_0}\int_0^{z_{source}}\,\frac{n_e(z)}{(1+z)^2\sqrt{\Omega_M(1+z)^3+\Omega_\Lambda}}\,dz\;,
\label{eq:y3}
\end{equation}
which is the formula in \cite{Macquart} where $\Lambda$CDM is assumed i.e. $H(z)\,=\,H_0\,\sqrt{\Omega_M(1+z)^3+\Omega_\Lambda}$. The number density of free electrons $n_e$ may be related to the number densities of hydrogen and helium atoms $n_{H}$, $n_{He}$ by $n_{e}(z)\,=\,X_H(z)\,n_{H}(z)+X_{He}(z)\,n_{He}(z)$ where $X_{H}$ and $X_{He}$ are the fractions of the ionized hydrogen and helium atoms, respectively. We have $n_{H}=\frac{f_{IGM}}{m_pc^2}\rho_H=\frac{f_{IGM}Y_H}{m_pc^2}\rho_b$, $n_{He}=\frac{f_{IGM}}{4m_pc^2}\rho_{He}=\frac{f_{IGM}Y_{He}}{4m_pc^2}\rho_b$ where $Y_{H}=\frac{3}{4}$, $Y_{He}=\frac{1}{4}$ are the baryonic mass fractions of hydrogen and helium, $f_{IGM}\simeq\,0.83$ is the fraction of baryons in intergalactic medium. Moreover, $\rho_b=\Omega_b\rho_0(1+z)^3=\frac{3\Omega_b\bar{H}_0^2}{8\pi\,G}(1+z)^3$. Hence, $n_{e}(z)\,=\,\frac{3\,f_{IGM}\,\Omega_b\bar{H}_0^2}{8\pi\,G\,m_pc^2}\left(X_H(z)Y_H\,+X_{He}(z)Y_{He}\,\right)(1+z)^3$, so (\ref{eq:y3}) becomes \cite{Wei}
\begin{equation}
DM_{IGM}
\,=\,\frac{3\,\Omega_b\,f_{IGM}\,\bar{H}_0^2}{8\pi\,G\,m_pc\,H_0}\int_0^{z_{source}}\,\frac{(1+z)\,\chi_e}{\sqrt{\Omega_M(1+z)^3+\Omega_\Lambda}}\,dz\;,
\label{eq:y4}
\end{equation}
where $\chi_e\,=\,X_H(z)Y_H\,+X_{He}(z)Y_{He}$.

The total dispersion measure $DM_{obs}$ is determined from time of arrival of FRB signals (which, in turn, may be determined from the line-of-sight of the group velocity) \cite{FRB2}. $DM_{MW}$ may be estimated \cite{DM-MW}. Hence, $DM_{h}+DM_{IGM}$ may be determined. Finally, this information and (\ref{eq:y4}) with the parameters except the Hubble constant being taken as the parameters found in Planck results based on Big Bang nucleosynthesis together with use of some statistical methods and simulations may be employed to find a best-fit value for the Hubble constant \cite{Wei,Gao}. These are the main lines of the procedure to determine Hubble constant from FRB observations. Note that \cite{Wei,Gao} and other studies for determining the Hubble constant, by use of FRB data, take $\bar{H}_0=H_0$ while we take $\bar{H}_0\neq\,H_0$ in general by Eq.(\ref{eq:49daa}).

After comparing (\ref{eq:y4}) with the standard procedure (where there is no distinction between $H_0$ and $\bar{H}_0$), it is observed that the effective Hubble constant measured in FRB measurements is, in fact, $\hat{H}_0=\left(\frac{\bar{H}_0}{H_0}\right)\bar{H}_0$ rather than $H_0$ or $\bar{H}_0$. In other words,
\begin{equation}
\hat{H}_0=\left(\frac{\bar{H}_0}{H_0}\right)\bar{H}_0\,<\,\bar{H}_0\,<\,H_0\;,  \label{eq:y5}
\end{equation}
where Eq.(\ref{eq:49daa}) is employed. 

Eq.(\ref{eq:y5}) is essentially the same as Eq.(\ref{eq:58aa}). (\ref{eq:y5}) and (\ref{eq:58aa}) together imply that the values of the Hubble constant obtained in Planck, BAO, and FRB measurements should be the same, and may be taken as $\hat{H}_0\,=\,\left(67.4\pm\,0.5\right)km\,s^{-1}\,Mpc^{-1}$. Moreover, Eq.(\ref{eq:y5}) tells us that $\hat{H}_0$ is smaller than $\bar{H}_0$ (that is the value Hubble constant in the parameterization of energy densities) which, in turn, is smaller than the value of the Hubble constant obtained in direct measurements (such as SN Ia measurements) and may be taken as the value measured by SHOES $H_0\,=\,\left(73.04\pm\,1.04\right)km\,s^{-1}\,Mpc^{-1}$. These values and (\ref{eq:y5}) may be used to predict the value of $\bar{H}_0$
\begin{equation}
\bar{H}_0\,=\,\sqrt{\hat{H}_0H_0}
\,=\, \sqrt{\left(67.4\pm\,0.5\right)\left({73.04\pm\,1.04}\right)}\;km\,s^{-1}\,Mpc^{-1}
\,\simeq\,70.16\pm\,0.56\;km\,s^{-1}\,Mpc^{-1},
\label{eq:y5a}
\end{equation}
The value of $\bar{H}_0$ is calculated from energy densities in \cite{Percival2}
\begin{equation}
\bar{H}_0\,=\,69.0\pm\,2.5\;km\,s^{-1}\,Mpc^{-1},
\label{eq:y5ax}
\end{equation}
Although this value is denoted by $H_0$ in \cite{Percival2}, in fact, it corresponds to  $\bar{H}_0$ since $G$ (rather than $G_{eff}$) is used in this study in the the expression for the critical energy density. The values in (\ref{eq:y5a}) and (\ref{eq:y5ax}) are compatible, although a more precise observational value of $\bar{H}_0$ (derived from energy densities) is needed for reaching a final conclusion on applicability of (\ref{eq:y5}). 

As already have been mentioned above, this scheme predicts that the value of Hubble constant measured by Planck $\left(67.4\pm\,0.5\right)km\,s^{-1}\,Mpc^{-1}$ should be the same as the one measured in FRB measurements. The values of the Hubble constant measured in FRB measurements (in terms of are $km\,s^{-1}\,Mpc^{-1}$) are $62.3\pm\,9.1$ \cite{Hagstotz}, $68.81^{+4.99}_{-4.33}$ \cite{Wu}, $71\pm\,3$ \cite{Liu}, $73.81^{+13}_{-8}$ \cite{DM-MW}, $80.4^{+24.1}_{-19.4}$ \cite{Zhao}, $95.8^{+7.8}_{-9.2}$ \cite{Wei}, $64^{+15}_{-13}$ \cite{Hoffmann}, $74^{+7.5}_{-7.2}$ \cite{Yang}, $70.41^{+2.28}_{-2.34}$ \cite{Gao}, $69.04^{+2.30}_{-2.07}$ (and $75.61^{+2.23}_{-2.07}$ assuming galactic electron density models) \cite{Wang}, (For FRB datasets with confirmed DM-z relation: $65.13\pm\,2.52$ for maximum likelihood, $57.67\pm\,11.99$ for arithmetic mean, $51.27^{+3.80}_{-3.31}$ for a linear DM-z relation, $77.09^{+8.89}_{-7.64}$ for a power-law DM-z relation) while for all FRB data sets: $\hat{H}_{0;Like}=67.30\pm\,0.91$, $\hat{H}_{0;Mean}=66.21\pm\,3.46$, The mean value for all FRB data sets in \cite{Piratova-Moreno} $\hat{H}_{0;Median}=66.10\pm\,1.89$, $\hat{H}_{0;Linear}=54.34\pm\,1.57$, $\hat{H}_{0;Power-law}=91.84\pm\,1.82$) \cite{Piratova-Moreno}; respectively. 
These values are in agreement with the Planck's value at 1 $\sigma$ level except those in \cite{Liu}, \cite{Wei}, \cite{Gao}, \cite{Wang} in the case of assuming galactic electron density models, \cite{Piratova-Moreno} in the case of assuming linear DM-z relation and power-law DM-z relation for datasets with confirmed DM-z relation and for $\hat{H}_{0;Linear}$ and $\hat{H}_{0;Power-law}$. The FRB values \textcolor{blue}{are} compatible with those of Planck at 2 $\sigma$ level except for $\hat{H}_{0;Linear}$ and $\hat{H}_{0;Power-law}$ mentioned above.

\section{Resolution of the Hubble tension by gravitational vacuum polarization}

The result of the previous section may be summarized as follows. It had been shown in \cite{Erdem-Universe} that the effect of inclusion of GPP (to be more preccise, the effect of GVP) is to raise the effective value of the Hubble parameter in Friedmann equation by multiplying it by an overall constant (which may be considered as multiplication of the Newton's gravitational constant by an overall constant that is geater than one). Therefore, GVP increases the value of the Hubble constant $H_0$ obtained in its direct measurements (i.e. directly from expansion \textcolor{blue}{rate} of the universe). On the other hand, as is evident from (\ref{s8-eq:0}), energy densities are not affected by GVP, so the value of the Hubble constant used in the parametrization of energy densities $\bar{H}_0$, is not affected by GVP, hence remains the same. As we will see below the difference between $H_0$ and $\bar{H}_0$ may be wholly attributed to GVP if the value of  $\sum_i\left(\frac{m_i\,c^2}{eV}\right)$ is taken accordingly. 

The value of the Hubble constant obtained from CMB observations (such as the Planck measurement) and FRB measurements i.e. $\hat{H}_0$ is related to $H_0$ and $\bar{H}_0$ by (\ref{eq:58aa}) and (\ref{eq:y5}) as $\tilde{H}_0=\left(\frac{\bar{H}_0}{H_0}\right)\bar{H}_0$.  It should be remarked that no new physics is employed in this approach. The model employed in this context (both in direct and indirect measurements) is just $\Lambda$CDM (both at the background and at the level of perturbations) since the effect of GPP just amounts to multiplying the Newton constant by an overall constant as is evident from Eq.(\ref{eq:49d}). Therefore, no new data analysis (in addition to that of $\Lambda$CDM) is needed for CMB and BAO data sets (unlike the extensions of $\Lambda$CDM model \cite{why}). The values obtained from these data sets (with $\Lambda$CDM adopted) remain applicable.

The equations (\ref{eq:49daa}), (\ref{s8-eq:2a}), (\ref{eq:58aa}), and (\ref{eq:y5}) imply
	\begin{equation} \left(\frac{\bar{H}_0}{H_0}\right)=\left(\frac{\hat{H}_0}{\bar{H}_0}\right)\,=\,\left[\,1-\,1.8\,\times\,10^{-58}\times\,\sum_i\left(\frac{m_i\,c^2}{eV}\right)^2\right] \label{n1}
	\end{equation}
	Eq.(\ref{n1}) suggests that the effect of GVP may be considerable for significant values of $\sum_i\left(\frac{m_i\,c^2}{eV}\right)^2$. For example, two $\phi$ particles each with a mass of the Planck mass contribute to $H_0$ so that $\bar{H}_0$ is multiplied approximately by 1.027. Hence, if the $\bar{H}_0$ given in (\ref{eq:y5ax}) is adopted, then $H_0$ and $\hat{H}_0$ are obtained as $H_0\,=\,70.9\,\pm\,2.6$ and $\hat{H}_0\,=\,67.2\,\pm\,2.4$, respectively, which are compatible with SN Ia and CMB, FRB values mentioned above. The $H_0$ and $\hat{H}_0$ values may be modified by modifying the value of $\sum_i\left(\frac{m_i\,c^2}{eV}\right)^2$. Increasing $\sum_i\left(\frac{m_i\,c^2}{eV}\right)^2$ increases $H_0$ while decreasing $\hat{H}_0$. In the next section we will consider the effect of GVP on another cosmological tension, namely, the $\sigma_8$ tension.

\section{Gravitational vacuum polarization and the $\sigma_8$ tension}

\subsection{Impact of gravitational vacuum polarization on the $\sigma_8$ tension}

The $\sigma_8$ tension which is the discrepancy between different measurements of the present time value of the matter density fluctuation amplitude (i.e the root-mean-square matter density matter perturbations) on scales of 8 $h^{-1}$ Mpc at present time. To be more specific, the $\sigma_8$ tension is the discrepancy between the values of $\sigma_8$ that are obtained directly at present time (e.g. from galaxy distrubution or weak lensing) and the values of $\sigma_8$ that are obtained by projecting the amplitude of matter perturbations obtained from CMB measurements (such as those of Planck) to the present time by using the evolution equations for perturbations.

The amplitude of matter density perturbations at present at scale $R$ is defined as
\begin{equation}
	\sigma_R^2\,\equiv\,<\left[\delta_M(\vec{r},z)\right]^2>_R\;,
	\label{s8-eq:1}
\end{equation}
where $\delta_M=\frac{\delta\,\rho_M}{\bar{\rho}_M}$ is the density contrast of matter with $\bar{\rho}_M$ being the average energy density of the background matter.
To be more specific
\begin{equation}
	\sigma_R^2(z)\,\equiv\,<\left[\delta_M(\vec{r},z)\right]^2>_R\,=\,\frac{1}{2\pi^2}\int\,dk\,k^2\,P\left(k,z\right)\,W^2\left(kR\right) \;,
	\label{s8-eq:4}
\end{equation}
where $P\left(k,z\right)$ is the power spectrum and
\begin{equation}
	W\left(kR\right)\,=\,\frac{3}{k^2R^2}\left(\frac{\sin(kR)}{kR}\,-\, \cos(kR\right)
	\label{s8-eq:5}
\end{equation}
is the Window function.

$\sigma_8$, in particular, is the amplitude of matter density perturbations at the scale $R\,=\,8\,h^{-1}$ Mpc at present time.
$\sigma_R$ in (\ref{s8-eq:1}) (so, $\sigma_8$) is either directly obtained from two-point correlation functions of galaxies, weak lensing etc. or by evolving the perturbations obtained from CMB (such as Planck observations) by use of the evolution equations for perturbations. There is a discrepancy between the directly and the indirectly obtained values of $\sigma_8$. For example, as mentioned in Introduction, there is a $2.6\sigma$ tension between KiDS-1000's directly obtained (obtained from using gravitational lensing) value and the Planck's indirectly determined value. This is called the $\sigma_8$ tension. In the following paragraphs we will see the impact of gravitational particle production (GPP), or to be more specific, the impact of gravitational vacuum polarization (GVP) on the $\sigma_8$ tension.

In principle GVP may have some impact on the evolution equations for perturbations. To linear order, at large scales, the matter density perturbation contrast $\delta_M$ satisfies \cite{growth1,growth2,growth3}
\begin{equation}
	\ddot{\delta}_M\,+\,2H\,\dot{\delta}_M\,-\,4\pi\,G_{eff}\,\bar{\rho}_M\,\delta_M \,=\,0 \;, \label{s8-eq:2}
\end{equation}
where $G_{eff}$ is defined by (\ref{s8-eq:2a}), and $H\,=\,\frac{\dot{a}}{a}$ is the value of the Hubble parameter in the Friedmann equation (\ref{eq:49d}) (i.e. the value of the Hubble parameter determined in its direct measurements).
}

Eq.(\ref{s8-eq:2}) may be re-expressed as
\begin{equation}
\frac{df}{d\ln{a}}\,+\,f^2\,+\,\left(\frac{\dot{H}}{H^2}+2\right)\,f\,-\,4\pi\,\frac{G_{eff}}{H^2}\,\bar{\rho}\,\textcolor{blue}{=\,0}\,\;, \label{s8-eq:3}
\end{equation}
where $f\,=\,\frac{d\ln{\delta_M}}{d\ln{a}}$.

It is evident that (\ref{s8-eq:3}) remains the same under a re-scaling of the Hubble parameter by Eq.(\ref{eq:49da}) where $G_{eff}$ is re-scaled as in  (\ref{s8-eq:2a}). This implies that the Hubble parameter $H$ and the effective Newton constant $G_{eff}$ in (\ref{s8-eq:3}) may be replaced by $\bar{H}$ and $G$ without changing $f(a)$ i.e. Eq.(\ref{s8-eq:3}) is equivalent to
\begin{equation}
\frac{df}{d\ln{a}}\,+\,f^2\,+\,\left(\frac{\dot{H}}{H^2}+2\right)\,f\,-\,4\pi\,\frac{G}{\bar{H}^2}\,\bar{\rho}\,\textcolor{blue}{=\,0}. \label{s8-eq:3x}
\end{equation}
In other words, the $f(a)$ obtained after the inclusion of the effect of GVP is the same as that would be obtained by Planck data (without inclusion of GVP). This, in turn, implies that inclusion of the effect of GVP does not have any impact on the $\sigma_8$ tension in the linear scale i.e. the value of $\sigma_8$ after the including the effect of GVP is the same as that the one without including the effect of GVP. This, in turn, implies that, at linear scales, the $\sigma_8$ value obtained from the Planck measurements after inclusion of the effect of GVP is the same as that the value obtained by Planck (without inclusion of  the effect of GVP).

A comment is in order here. Although we have adopted the $G$, $\bar{H}^2$ pair in (\ref{s8-eq:3x}) since the energy densities are not affected by GVP and $G$ is employed in the perturbation evolution equations by default. However, $\frac{G_{eff}}{H^2}$ in (\ref{s8-eq:3}) may be replaced by any set of gravitational constant and Hubble parameter with the same ratio as $\frac{G_{eff}}{H^2}$ (e.g. $\frac{G}{\bar{H}^2}$). In fact, this is true for all perturbation evolution equations. The cosmological perturbations remain the same under the transformation
$G\,\rightarrow\,\lambda^2\,G$, $\bar{H}^2\,\rightarrow\,\lambda^2\,\bar{H}^2$ except at the times when matter - gravity interactions dominate their evolution (such as in the interplay between Thomson scattering and the expansion rate of the universe at the time of recombination \cite{Zahn,Lamine}). The same is true for the fundamental Friedmann equation  Eq.(\ref{eq:49d}). We may multiply both sides of the equation by the same constant (i.e. multiply the Hubble constant and the Newton's constant by the same constant without changing its mathematical content. Out of these possible infinitely many Hubble constant we choose one of them i.e. $H_0$ as the one that is directly measured by the expansion rate of the universe as the physically relevant value. In the case of CMB, any set of gravitational constant and Hubble parameter with the same ratio as $\frac{G}{\bar{H}_0^2}$ may be adopted as the relevant set in the absence of information from recombination or BBN. However recombination and BBN break this degeneracy since the redshifts of recombination and BBN are determined by a balance of the rates of particle physics processes (such as Thomson scattering) and the expansion rate of the universe (i.e. H). This, in turn, implies that one gets the $G$, $\bar{H}_0^2$ pair if the gravitational constant is imposed to be $G$. On the other hand, one will get the $G_{eff}$, $H_0^2$ pair when there is no constraint on the value of the gravitational constant and information from recombination or BBN eras is provided since inclusion of information from the era of recombination imposes the $G_{eff}$, $H_0^2$  pair out of infinitely many pairs with the ratio $\frac{G}{\bar{H}_0^2}$. In fact this prediction seems to be verified in \cite{Lamine} for some sets of data sets. Its data analysis results in the Hubble constant values close that of Planck when the gravitational constant is taken as $G\,=\,G_N\,=\,6.67430\,\times\,10^{-11}m^3kg^{-1}s^{-2}$ \cite{CODATA} i.e. the value of the Newton's gravitational in the absence of gravitational vacuum polarization (which may be identified as the Newton's gravitational constant measured on Earth by torsion balance and pendulum experiments \cite{measurement-G-Earth}). On the other hand, \cite{Lamine} finds the set $G_{eff}$, $H_0^2$ (where $H_0$ is approximately equal to the value of the Hubble constant obtained by SN Ia measurements) for some other data sets. For example, it finds the Hubble constant as $H_0\,=\,74.4\,^{+3.7}_{-5.2}\,km\,s^{-1}\,Mpc^{-1}$, $G_{eff}\,=\,\left(1.048^{0.027}_{-0.035}\right)\times\,G_N$ for P20 TT data set \cite{P20TT} while some other data sets result in values close to $G$, $\bar{H}_0^2$ . On the other hand, \cite{measurement-G-cosmological-1} finds values of the Hubble constant and the gravitational constant close to $\bar{H}_0^2$ , $G$. Elucidating the source of this discrepancy (in the values of the Hubble and the Newton's gravitational constant) between different data sets (when the Newton's gravitational constant is treated as a free parameter) may be useful.

Another point worth to mention is that it is possible that the evolution equations in the non-linear scales do not remain invariant under this re-scaling.  However, studies \cite{Ma,Sanchez2}, show that different models with the same evolution for perturbations at linear scales have approximately the same power spectra at mildly non-linear scales. Similar studies by use of halo mass functions result in similar conclusions \cite{Gilman}. Although, there are some studies raising the possibility of resolving the Hubble tension by a possible suppression of matter power spectrum at non-linear scales \cite{Preston}, the observational results are not conclusive regarding this possibility \cite{Shah}. Moreover, the window function $W\left(kR\right)$ in (\ref{s8-eq:5}) smooths the contribution of non-linearities in the matter power spectrum. Therefore, I will assume that the evolution of perturbations in the linear regime approximately survives even after inclusion of the effects of non-linear contributions (such as baryon feedback. Hence, the comments about the impact of GVP on the $\sigma_8$ tension mentioned in the preceding paragraph for the linear regime (i.e. GVP does not neither worsens nor improves $\sigma_8$ tension) survives (at least approximately) even after including the effect of non-linear scales on $\sigma_8$.

\subsection{Contribution of the ambiguity of the definition of $\sigma_8$ to the $\sigma_8$ tension}

In the preceding subsection we have argued that the matter density perturbations at the scale $R\,=\,8\,h^{-1}$ Mpc at present are not significantly affected 
by GVP. Therefore, the status of the $\sigma_8$ tension after the effect of GVP is included is the same as that of $\Lambda$CDM before the effect of GVP included. At first sight this seems to suggest that inclusion of the effect of GVP does make any progress in the resolution of the the $\sigma_8$ tension. Nonetheless, the present scheme (of the resolution of the Hubble tension by inclusion of the effect of GVP) has an important superiority over the other schemes since other schemes that try to remove or mitigate the Hubble tension generically worsen the $\sigma_8$ tension \cite{worsen-s8-misuse} while this is not the case for the present scheme. This point is important in the light of \cite{Sanchez,worsen-s8-misuse}. \cite{Sanchez,worsen-s8-misuse} find that the source of the $\sigma_8$ tension may be the ambiguity introduced by the presence of $h$ in the definition of the scale of $\sigma_8$, $R\,=\,8\,h^{-1}$ Mpc in (\ref{s8-eq:5}). \cite{worsen-s8-misuse} derived the following formula that relates $\sigma_8$ to $h$ by
\begin{equation}
\sigma_8(z)\,\approx\,\sigma_{12}(z)\,\left[1\,+\,1.035\,\delta\,h \right]\,+\,{\cal O}(\delta\,h\,^2)\;,
\label{s8-eq:4x}
\end{equation}
where
$\sigma_{12}(z)$ is the value of $\sigma_R(z)$ at $R\,=\,12$ Mpc (i.e. $\sigma_8(z)$ for $h\,=\,\frac{8}{12}$), $\delta\,h\,=\,h-h_*$ with $h_*=\frac{8}{12}\simeq\,0.667$. It is evident that taking this point into account mitigates the $\sigma_8$ tension. For example if one lets $h\,=\,\frac{9}{12}$ then one obtains $\sigma_8(0)\,\approx\,\left[1\,+\,\frac{1.035}{12}\,\right]\sigma_{12}(0)\simeq\,\frac{13}{12}\sigma_{12}(0)$ i.e. $\sigma_8\,=\,0.834$ of Planck (for $h\,\sim\,\frac{8}{12}$) is $\frac{13}{12}$ times of the value of $\sigma_8\,=\,0.759$ of KiDS-1000 (for $h\,\sim\,\frac{9}{12}$) which is really approximately satisfied.

In other words, the situation of the $\sigma_8$ tension is the same as that of $\Lambda$CDM after the effect of GVP is included while the models that are proposed to solve the Hubble tension generically worsen the $\sigma_8$ tension. Moreover, given some studies that find no indication of the tension \cite{no-tension}, the $\sigma_8$ tension, at least partially, may be attributed to the ambiguity of the  definition of $\sigma_8$ as mentioned above, the situation of the Hubble and $\sigma_8$ tensions after inclusion of the effecr of GVP is better than most of the models proposed for the solutions of these tensions \cite{why,worsen,simultaneous,visual}. However, note that, as we have mentioned in Introduction, there is still a possibility to have an $S_8$ tension induced by a possible tension between the values of $\Omega_m$ measured in different observations \cite{DESI1,DESI2}.

\section{Conclusion}

 In \cite{Erdem-Universe}, it had been shown that inclusion of the effect of gravitational vacuum polarization (GVP) into account either may relieve or mitigate the Hubble tension depending on the total mass of the scalars in the spectrum of the relevant particle physics model. To this end, in this manuscript, first we have given a brief overview of \cite{Erdem-Universe} to remind that GVP increases the effective value of the Hubble constant $H_0$ (i.e. the true Hubble constant) in its direct measurements while it does not change the value of the Hubble constant $\bar{H}_0$ in the parametrization of energy densities. A misidentification in \cite{Erdem-Universe}, namely, identification of the Hubble constant in CMB and BAO measurements as $\bar{H}_0$, is also corrected in the present paper; the Hubble constant in CMB and BAO measurements is, in fact, $\hat{H}_0$.  
 
  In the light of the observation that the expression for the dispersion measure for intergalactic medium $DM_{IGM}$ depends both on true Hubble parameter $H\,=\,\frac{\dot{a}}{a}$ and energy density of baryons (that is parameterized in terms of $\bar{H}$), we have found the same expression for the Hubble constant in FRB measurements as the expression in the case of CMB and BAO measurements, namely, $\hat{H}_0=\left(\frac{\bar{H}_0}{H_0}\right)\bar{H}_0$. In principle, this prediction may be checked by using the observational values of $\hat{H}_0$, $H_0$, $\bar{H}_0$. Confirmation of $\hat{H}_0=\left(\frac{\bar{H}_0}{H_0}\right)\bar{H}_0$ by observational data in future would imply that gravitational vacuum polarization is the only source of the Hubble tension.  A quick check on the degree of the compatibility between the prediction of this equation and observational data at 1 $\sigma$ level suggests that that the precision of the present observational results are not conclusive enough yet. More precise observations and more detailed analysis in future may result in a more definite picture on this point.

 Then, we have considered the effect of GVP on the $\sigma_8$ tension. It has been observed that re-scaling of Hubble parameter does not change the (linear) evolution equations for matter perturbations. Such a re-scaling is equivalent to replacing the true Hubble parameter $H$ by $\bar{H}$ while replacing the effective value of Newton's constant $G_{eff}$ (that includes the effect of GVP) by Newton's constant $G$ (without the effect of GVP). This, in turn, is equivalent to taking the Hubble constant in the linear evolution equations of matter perturbations as $\bar{h}\,\sim\,0.67$, so the resulting $\sigma_8$ is that of Planck. Therefore, the situation after inclusion of the effect of GVP is the same as that of $\Lambda$CDM (at linear scales). This is the superiority of explaining the Hubble tension by the effect of GVP since the models that are proposed for resolution of the Hubble tension generically worsen the $\sigma_8$ tension \cite{worsen-s8-misuse} while the magnitude of the $\sigma_8$ after inclusion of the effect of GPP is the same as that of $\Lambda$CDM (before inclusion of the effect of GVP). Moreover, \cite{Sanchez,worsen-s8-misuse} have shown that the ambiguous definition of $\sigma_8$ in terms of the Hubble constant may be the main source of the the $\sigma_8$ tension (or there may be no $\sigma_8$ tension in $\Lambda$CDM at all). Furthermore,  by studying observational results for cluster mass functions and some popular theoretical halo mass functions in the context of $\Lambda$CDM, \cite{no-tension} finds no indication for presence of a $\sigma_8$ tension. (However, a precise conclusion needs study of the effect of GVP at non-linear scales and clarification of the issue of baryonic feedback mechanisms \cite{nonlinear}. This point requires a separate study by its own.) In the light of these considerations, GVP may provide insight into both of the Hubble and  the $\sigma_8$ tensions.

Finally, it is useful to note that Eq.(\ref{eq:49d}) holds for all models of Friedmann-Lema{\^{i}}tre-Robertson-Walker cosmologies (although $\Lambda$CDM is assumed in the context of CMB and FRB measurements at present). At most, an additional condition that it should mimic $\Lambda$CDM  may be imposed (so that the adiabatic conditions are ensured \cite{Erdem-Universe}).  If the true cosmological model turns out to be different from $\Lambda$CDM (as (\cite{running-H0-2}) suggests), then the expressions for $H(z)$ in CMB and FRB measurements must be modified accordingly. In that case the data in different redshift bins would result in different Hubble constants if the cosmological model is assumed to be  $\Lambda$CDM  \cite{running-H0-1,running-H0-2}. However, there would still be three different kinds of Hubble constants ($H_0$, $\bar{H}_0$, $\hat{H}_0$). Out of these $H_0$ is the true Hubble constant in the sense that only $H_0$ directly measures the expansion rate of the rate at $z=0$. If use of $\Lambda$CDM is insisted while the true cosmological model differs from $\Lambda$CDM,  then each of $\bar{H}_0$ and $\hat{H}_0$ will have different Hubble constants in different redshift bins. $H_0$ has a single value if it is assumed to be directly measured at $z=0$ through expansion rate of the universe. However, if $H_0$ is measured at different $z\sim\,0$, then the resulting values will be the same as those of $\bar{H}_0$ multiplied by $\sqrt{\frac{G_{eff}}{G}}$ at the same redshift values.

\appendix

\section{Impact of gravitational particle production on Hubble constant}

\subsection{Gravitational particle production: preliminaries}

Spacetime at cosmological scales may be described by spatially flat Robertson-Walker metric
\begin{equation}
ds^2\,=\,-dt^2\,+\,a^2(t)\left[dr^2+r^2
(d\theta^2+\sin^2{\theta}d\phi^2)\right]\;. \label{e1}
\end{equation}

We consider the following action for a scalar field $\phi$ in this space
\begin{eqnarray}
&S&
\,=\,\int\sqrt{-g}\;d^4x\,\frac{1}{2}\left[-g^{\mu\nu}\partial_\mu\phi\partial_\nu\phi\,-\,m_\phi^2\phi^2\right] \label{e2a} \,=\,
\int\,d^3x\,d\eta\,\frac{1}{2}\left[
\tilde{\phi}^{\prime\;2}-(\vec{\nabla}\tilde{\phi})^2
-\tilde{m}_\phi^2\tilde{\phi}^2\right]\;,
\label{e2b}
\end{eqnarray}
where $m_\phi$ is the mass of $\phi$, prime denotes derivative with respect to conformal time $\eta$ \cite{QFTC1} (while an over-dot denotes the derivative with respect to $t$) and
\begin{equation}
d\eta\,=\,\frac{dt}{a(t)}~,
~\tilde{\phi}\,=\,a(\eta)\;\phi~,~~~
\tilde{m}_\phi^2\,=\,m_\phi^2a^2-\frac{a^{\prime\prime}}{a}.
\label{e2aa}
\end{equation}

The field $\tilde{\phi}$ may be expressed as
\begin{equation}
\tilde{\phi}(\vec{x},\eta)\,=\,\frac{1}{\sqrt{2}}\,\int\,\frac{d^3k}{(2\pi)^\frac{3}{2}}\,\left[ e^{i\vec{k}.\vec{x}}\,v_k^*(\eta)\,\hat{a}_{\vec{k}}^-
\,+\, e^{-i\vec{k}.\vec{x}}\,v_k(\eta)\,\hat{a}_{\vec{k}}^+ \right]\;. \label{eq:3}
\end{equation}
The mode function $v_k(\eta)$ satisfies the equation of motion for $\tilde{\phi}$
\begin{equation}
v_k^{\prime\prime}\,+\,\omega_k^2\,v_k = 0\;,
\label{eq:4}
\end{equation}
where
\begin{equation}
\omega_k\,=\,\sqrt{\vec{k}^2\,+\,\tilde{m}_\phi^2} \;.
\label{eq:5}
\end{equation}

Vacuum state is the ground state with minimum energy. In curved space, in general, the ground state at an instant of time is not the ground state at another time. Hence, the annihilation operators corresponding to the corresponding vacuum state at a given time do not destroy the vacuum state at another time. Therefore, the annihilation and creation operators and the mode functions at different times are different in general in curved spaces \cite{QFTC1}. The field $\tilde{\phi}$ in another vacuum (other than the one specified in (\ref{eq:3})) with annihilation and creation operators $\hat{b}_{\vec{k}}^-$ and $\hat{b}_{\vec{k}}^+$ may be expanded as
\begin{equation}
\tilde{\phi}(\vec{x},\eta)\,=\,\frac{1}{\sqrt{2}}\,\int\,\frac{d^3k}{(2\pi)^\frac{3}{2}}\,\left[ e^{i\vec{k}.\vec{x}}\,u_k^*(\eta)\,\hat{b}_{\vec{k}}^-
\,+\, e^{-i\vec{k}.\vec{x}}\,u_k(\eta)\,\hat{b}_{\vec{k}}^+ \right] \;,  \label{eq:6}
\end{equation}
where $u_k$ satisfies the same equation as (\ref{eq:4}) and is related to $v_k$ and $v_k^*$ by
\begin{equation}
u_k(\eta)\,=\,\alpha_k\,v_k(\eta)\,+\,\beta_k\,v_k^*(\eta) \;, \label{eq:7}
\end{equation}
where $\alpha_k$, $\beta_k$ are called Bogolyubov coefficients. In similar fashion, $\hat{b}_{\vec{k}}^-$ is related to $\hat{a}_{\vec{k}}^-$ and $\hat{a}_{\vec{k}}^+$ by
\begin{equation}
\hat{b}_{\vec{k}}^-\,=\,\alpha_k\,\hat{a}_{\vec{k}}^-\,-\,\beta_k\,\hat{a}_{\vec{k}}^+ ~,~~
\hat{b}_{\vec{k}}^+\,=\,\alpha_k^*\,\hat{a}_{\vec{k}}^+\,-\,\beta_k^*\,\hat{a}_{\vec{k}}^- \;.
\label{eq:7}
\end{equation}
Eq.(\ref{eq:7}) may be inverted to write
\begin{equation}
\hat{a}_{\vec{k}}^-\,=\,\alpha_k^*\,\hat{b}_{\vec{k}}^-\,+\,\beta_k\,\hat{b}_{\vec{k}}^+~,~~
\hat{a}_{\vec{k}}^+\,=\,\alpha_k\,\hat{b}_{\vec{k}}^+\,+\,\beta_k^*\,\hat{b}_{\vec{k}}^-\;.
\label{eq:7a}
\end{equation}

Let $\hat{a}_{\vec{k}}^-$ ($\hat{a}_{\vec{k}}^+$) denotes the annihilation (creation) operator of the $in$ state while $\hat{b}_{\vec{k}}^-$ ($\hat{b}_{\vec{k}}^+$) denotes the annihilation (creation) operator of the $out$ state, and $|_(b)0>$ the vacuum state of the out particles. The number operator for the $in$ particles with momenta $\vec{k}$ is $\hat{N}_{\vec{k}}^{(a)}$ = $\hat{a}_{\vec{k}}^+\hat{a}_{\vec{k}}^-$. After expressing $\hat{a}_{\vec{k}}^-$ and $\hat{a}_{\vec{k}}^+$ in terms of $\hat{b}_{\vec{k}}^-$ and $\hat{b}_{\vec{k}}^+$, one finds \cite{QFTC1}
\begin{equation}
<_{(b)}0|\hat{N}_{\vec{k}}^{(a)}|_{(b)}0>\,=\,|\beta_k|^2\,\delta^{(3)}(0).
\label{eq:5a}
\end{equation}
This implies that the number density of the "in" particles with momentum $\vec{k}$ produced from the vacuum of the $out$ particles in the $out$ region is given by $|\beta_k|^2$.

\subsection{Gravitational particle production after decoupling}

Mode functions may be expressed in a WKB-approximation-like form
\begin{equation}
v_k(\eta)\,=\,\frac{1}{\sqrt{W_k(\eta)}}\,\exp{\left[i\int_{\eta_0}^\eta\,W_k(\eta)\,d\eta \right]} \;. \label{eq:8}
\end{equation}
The requirement of (\ref{eq:8}) being a solution of (\ref{eq:4}) implies that $W_k(\eta)$ should satisfy
\begin{equation}
W_k^2\,=\,\omega_k^2\,-\, \frac{1}{2}\left[ \frac{W_k^{\prime\prime}}{W_k}-\frac{1}{2}\left(\frac{W_k^{\prime}}{W_k}\right)^2 \right]\,+\,\frac{i}{2}\,W_k^\prime \;.
 \label{eq:9}
\end{equation}
If the expansion of the universe is adiabatic i.e. if it is sufficiently slow, then $\left|\frac{\omega_k^\prime}{\omega_k^2}\right|$ and $\left|\frac{\omega_k^{\prime\prime}}{\omega_k^3}\right|$ are small. In this case $W_k^2\,\approx\,\omega_k^2$, so $v_k(\eta)$  may be approximated
by \cite{Ford,Boyanovsky}
\begin{equation}
v_k(\eta)\,=\,\frac{1}{\sqrt{\omega_k(\eta)}}\,\exp{\left[i\int_{\eta_0}^\eta\,\omega_k(\eta)\,d\eta \right]} \;, \label{eq:31}
\end{equation}
where
\begin{equation}
\omega_k\,=\,\frac{1}{\hbar}\sqrt{\hbar^2\vec{k}^2c^2\,+\,m_\phi^2c^4a^2-\frac{a^{\prime\prime}}{a}\hbar^2} \;.
 \label{eq:44a}
\end{equation}

In \cite{Erdem-Universe} it is shown that for $\Lambda$CDM (and for a model that has a similar cosmological evolution as $\Lambda$CDM) the following relation holds
\begin{equation}
\frac{\tilde{m}_\phi^\prime}{\tilde{m}_\phi^2}\,\ll\,1~~\mbox{and}~~~\frac{\tilde{m}_\phi^{\prime\prime}}{\tilde{m}_\phi^3}\,\ll\,1~~~~~~~~\mbox{provided that}~~~~m_\phi\,c^2\,\gg\,10^{-27}\,eV \;,
\label{eq:25}
\end{equation}
which, in turn, implies
\begin{equation}
\left|\frac{\omega_k^\prime}{\omega_k^2}\right|\,\ll\,1
~~~~ \mbox{and}~~~\,
\left|\frac{\omega_k^{\prime\prime}}{\omega_k^3}\right|\,\ll\,1
~~~\,
~~~~ \mbox{provided that}~~~\,  m_\phi\,c^2\,\gg\,10^{-27}\,eV
 \label{eq:27}
\end{equation}
for $a\,>\,10^{-4}$ (i.e. after decoupling). Note that the upper limit on the mass of $\phi$ in this equation, namely, $\frac{m_\phi\,c^2}{eV}\,\gg\,10^{-27}$ is satisfied by all standard dark matter candidates including ultra-light dark matter. It is evident that
(\ref{eq:27}) guarantees (\ref{eq:31}). Moreover, $\omega_k$ in this case may approximated by
\begin{equation}
\omega_k\,\simeq\,\frac{1}{\hbar}\sqrt{\hbar^2\vec{k}^2c^2\,+\,m_\phi^2c^4a^2} \;.
 \label{eq:44aa}
\end{equation}

In \cite{Erdem-Universe} it is shown that the following relation is satisfied in a sufficiently small time interval $\Delta\,\eta$
\begin{equation}
\left|\frac{\Delta\,\omega_k}{\omega_k}\right|\,=\,
\left|\frac{\Delta\,\eta\,\left(\frac{d\omega_k}{d\eta}\right)}{\omega_k}\right|\,\ll\,1.
\label{eq:34}
\end{equation}
Hence (\ref{eq:3}) may be expressed in its Minkowski form in the interval $\Delta\,\eta$, so the field may expanded as \cite{Erdem-Gultekin1,Erdem-Gultekin2}
\begin{eqnarray}
\tilde{\phi}_   {(i)}(\vec{x},\eta)\,\simeq\,
\int\,
\frac{d^3\tilde{p}}{(2\pi)^\frac{3}{2}\sqrt{2\omega_{p,(i)}}}\left[a_{p,(i)}^-\,
e^{i\left(\vec{\tilde{p}}.\vec{r}-\omega_{p,(i)}(\eta-\eta_i)\right)}
\,+\,a_{p,(i)}^+
\,e^{i\left(-\vec{\tilde{p}}.\vec{r}+\omega_{p,(i)}(\eta-\eta_i)\right)}\right]\label{eq:35} \\
\eta_i\,<\,\eta\,<\,\eta_{i+1}\;, \nonumber
\end{eqnarray}
where $_{(i)}$ refers to the $i$th time interval between the times $\eta_i$ and $\eta_{i+1}$ with $\Delta\,\eta\,=\,\eta_{i+1}-\eta_i$.

A mode function corresponding to a ground state with minimum energy at time $\eta_0$ has the form $v_k(\eta_0)\,=\,\frac{1}{\sqrt{\omega_k\left(\eta_0\right))}}\,e^{i\sigma_k\left(\eta_0\right)}$ where $\sigma_k$ is an arbitrary function of $|\vec{k}|$ and $\eta_0$ \cite{QFTC1}. In the light of this and the above observations, the mode functions of out states, for example, may be expressed as mode functions of Minkowski space in each interval $\Delta\eta$ that satisfies (\ref{eq:34}) i.e.
\begin{equation}
v_k^{(out)}(\eta)\,=\,\frac{1}{\sqrt{\omega_k\left(\eta_f\right))}}\,e^{i\left[\omega_k\left(\eta_f\right)\,\eta\right]} \;, \label{eq:37}
\end{equation}
where
\begin{equation}
\omega_k\left(\eta_f\right)\,=\,c\,\sqrt{\vec{k}^2\,+\,\left(\frac{m_\phi\,c}{\hbar}\right)^2\,a_f^2}
  \label{eq:37a}
\end{equation}
 is the value of $\omega_k$ at a final time $\eta_f$, and $\eta_f-\Delta\eta\,<\,\eta\,<\,\eta_f$ with $\Delta\eta$ being sufficiently small so that (\ref{eq:34}) holds while not being extremely small. The mode functions of the in states may expressed in the same form as (\ref{eq:37}) where $\eta_f$ is replaced by initial time $\eta_i$, and the in states evolve at later times as in (\ref{eq:31}). (To be precise we identify  $\eta_i$ and $\eta_f$ as $a_i\,=\,a(\eta_i)\,\simeq\;~10^{-4}~-~10^{-3}$, $a_f\,=\,a(\eta_f)\,\simeq\;~1$.)

Now we use the matching conditions for the mode functions and their derivatives at boundaries at the time $\eta$ with $\eta_f-\Delta\eta\,<\,\eta\,<\,\eta_f$, namely,
\begin{eqnarray}
v_k^{(in)}(\eta)&=& \alpha_k\,v_k^{(out)}(\eta)\,+\,\beta_k\,v_k^{(out)*}(\eta)\,  \label{eq:38a} \;,\\
v_k^{(in)\prime}(\eta)&=& \alpha_k\,v_k^{(out)\prime}(\eta)\,+\,\beta_k\,v_k^{(out)*\prime}(\eta)\,  \label{eq:38b}
\end{eqnarray}
to determine $\beta_k$ (since $\left|\beta_k\right|^2$ is the number density of the gravitationally produced particles with momentum $\vec{k}$). Note that $v_k^{(in)}(\eta)$ in \ref{eq:38a}) and (\ref{eq:38b}) is its value in the $out$ region. In this region the form of $v_k^{(in)}(\eta)$ is given by (\ref{eq:31}) while that of $v_k^{(out)}(\eta)$ is given by (\ref{eq:37}).

By (\ref{eq:37}),
\begin{eqnarray}
\alpha_k\,v_k^{(out)}(\eta)\,+\,\beta_k\,v_k^{(out)*}(\eta)\,&=& \frac{1}{\sqrt{\omega_k\left(\eta_f\right))}}\,
\left[\alpha_k\,e^{i\left[\omega_k\left(\eta_f\right)\,\eta\right]}  \,+\,\beta_k\, e^{-i\left[\omega_k\left(\eta_f\right)\,\eta\right]} \,\right] \;,
\label{eq:39a} \\
\alpha_k\,v_k^{(out)\prime}(\eta)\,+\,\beta_k\,v_k^{(out)*\prime}(\eta)\,&=&
i\sqrt{\omega_k\left(\eta_f\right)}\,
\left[\alpha_k\,e^{i\left[\omega_k\left(\eta_f\right)\,\eta\right]}  \,-\,\beta_k\, e^{-i\left[\omega_k\left(\eta_f\right)\,\eta\right]} \,\right]\;.
\label{eq:39b}
\end{eqnarray}
In the following we will let $\eta_f=\eta$ since we consider generic $\eta_f$ i.e. we may replace $\eta_f$ in (\ref{eq:39a}) and (\ref{eq:39b}) by $\eta$ provided that $\eta$ is in a sufficiently small interval $\Delta\eta$.

Hence, after using (\ref{eq:31}) for $v_k^{(in)}(\eta)$ and making use of (\ref{eq:38a}), (\ref{eq:38b}), (\ref{eq:39a}), (\ref{eq:39b}), we find
\begin{equation}
\left|\beta_k\right|^2
\,=\,\frac{\left(\frac{m_\phi\,c}{\hbar}\right)^4\,a^2\,a^{\prime\,2}}{16c^2\left[\vec{k}^2\,+\,\left(\frac{m_\phi\,c}{\hbar}\right)^2a^2\right]^3} \;.
\label{eq:47}
\end{equation}

The energy density corresponding to (\ref{eq:47}) is
\begin{equation}
\rho^{(PP)}\,=\, \frac{1}{(2\pi)^3\,a^3}\int\,d^3k\,E_k\,\left|\beta_k\right|^2
\,=\,\frac{\hbar}{96\,\pi\,c}\,
\left(\frac{m_\phi\,c}{\hbar}\right)^2H^2 \;,
\label{eq:48xx}
\end{equation}
where $E_k\,=\,\sqrt{\hbar^2\left(\frac{\vec{k}}{a}\right)^2c^2\,+\,m_\phi^2c^4}$ is the physical energy of a $\phi$ particle with physical momentum $\frac{1}{a}\hbar\vec{k}$ (while $\hbar\vec{k}$ is the comoving coordinate momentum of the particle).

\section{Gravitational vacuum polarization versus gravitational particle production}

In the discussion after Eq.(\ref{eq:49d}) we had argued that the result in that equation should be interpreted as some kind of a gravitational vacuum polarization. In this appendix we give the argument in a more formal way.

The formula (\ref{eq:47}), so (\ref{eq:48xx}) are obtained for the case where  (\ref{eq:27}) is satisfied i.e. when (\ref{eq:44aa}) holds. In the general case, (\ref{eq:38a}) together with (\ref{eq:38b}) and (\ref{eq:39a}) together with (\ref{eq:39b}) and (\ref{eq:31}) together with (\ref{eq:44a}) may be solved for $\beta_k$ as
\begin{equation}
|\beta_k|\,=\,\left|\frac{\omega_k^\prime}{4\omega_k^\frac{3}{2}}v_k^{(in)}\right| \;,
\label{eq:x1}
\end{equation}
which, in turn, results in,
\begin{equation}
|\beta_k|^2\,=\,\frac{\left[\left(\tilde{m}^2\right)^\prime\right]^2}{64\omega_k^6}
\,=\,
\frac{\left[2aa^\prime\,m_\phi^2\,-\,\left(\frac{a^{\prime\prime}}{a}\right)^\prime\right]^2}{64\omega_k^6} \;,
\label{eq:x2}
\end{equation}
where we have used (\ref{eq:5}) and (\ref{eq:31}).

In the special case of (\ref{eq:27}) i.e. when $\tilde{m}^2\,\simeq\,m^2a^2$, (\ref{eq:x2}) reduces to (\ref{eq:47}). The $2aa^\prime\,m_\phi^2$ term in (\ref{eq:x2}) is the term that results in Eq.(\ref{eq:48xx}). To see the content of the other term in (\ref{eq:x2}) i.e. the $-\,\left(\frac{a^{\prime\prime}}{a}\right)^\prime$ term more explicitly it is useful to consider the special case
\begin{equation}
H\,=\,\frac{\dot{a}}{a}\,=\,\zeta\,a^s \;,
\label{eq:x3}
\end{equation}
where $\zeta$ and $s$ are some constants. In fact, (\ref{eq:x3}) essentially includes all cosmologically interesting main eras, namely, the de Sitter, radiation dominated, matter dominated eras as subcases that correspond to $s\,=\,0$, $s\,=\,-2$, $s\,=\,-\frac{3}{2}$; respectively. In this case (\ref{eq:x2}) becomes
\begin{equation}
|\beta_k|^2\,=\,
\frac{a^6\,m_\phi^4\,H^2\,\left[1\,-\,(s+2)\left(\frac{H}{m_\phi}\right)^2\right]^2}{16\omega_k^6}\;.
\label{eq:x4}
\end{equation}

In the case $\frac{H}{m_\phi}\,\ll\,1$, (\ref{eq:x4}) reduces to (\ref{eq:47}). $\frac{H}{m_\phi}\,\ll\,1$ is really satisfied for (\ref{eq:27}) since $H_0\hbar\,\simeq\,1.5\,\times\,10^{-33}eV$ is much smaller than $10^{-27}eV$ so $\frac{H}{m_\phi}\,\ll\,1$ till the redshifts of order of $10^4$. However, in other cases in the cases where $\frac{H}{m_\phi}$ is not small (e.g. in inflationary era \cite{Starobinsky-versus-Bogolyubov}) the $(s+2)\left(\frac{H}{m_\phi}\right)^2$ may also have non-neglible contribution. This term introduces additional terms to $\rho^{(PP)}$ that are proportional to $H^4$ and $H^6$. However, these terms can not be included in the redefinition of $G$ in Friedmann equations as done in (\ref{eq:49d}). Therefore, the effect of gravitational particle production cannot be identified only by gravitational vacuum polarization when $\frac{H}{m_\phi}$ is non-negligible .

\end{document}